\documentclass[aps,prl,twocolumn,amsmath,amssymb,floatfix,superscriptaddress]{revtex4-2}
\usepackage{tabularx}
\usepackage{bm}
\usepackage{euscript}
\usepackage{epsfig,psfrag,subfigure}
\usepackage{graphicx}
\usepackage{color}
\usepackage{amsfonts}
\usepackage{exscale}
\usepackage{amsbsy}
\usepackage{ulem}
\usepackage{multirow}

\def\avg#1{\left\langle#1\right\rangle}
\def\bra#1{\left\langle#1\right|}
\def\ket#1{\left|#1\right\rangle}
\def\braket#1#2{\left\langle #1\right|\left.#2\right\rangle}

\def\abs#1{\left|#1\right|}

\def\sgn{{\rm sgn}}

\def\be{\begin{equation}}       \def\ee{\end{equation}}
\def\bea{\begin{eqnarray}}      \def\eea{\end{eqnarray}}
\def\ba{\begin{array} }
\def\ea{\end{array} }

\def\=>{\Rightarrow}
\def\->{\rightarrow}
\def\A{\uparrow}
\def\V{\downarrow}
\def\Eq#1{Eq.~(\ref{#1})}
\def\Fig#1{Fig.~\ref{#1}}

\renewcommand{\v}[1]{{\bf #1}}

\begin{document}

\title{Possible superconductivity with Bogoliubov Fermi surface in lightly doped Kagome U(1) spin liquid}
\author{Yi-Fan Jiang}
\affiliation{School of Physical Science and Technology, ShanghaiTech University, Shanghai 201210, China}
\affiliation{Stanford Institute for Materials and Energy Sciences, SLAC National Accelerator Laboratory and Stanford University, Menlo Park, CA 94025, USA}
\author{Hong Yao}
\email{yaohong@tsinghua.edu.cn}
\affiliation{Institute of Advanced Study, Tsinghua University, Beijing 100084, China}
\affiliation{State Key Laboratory of Low Dimensional Quantum Physics, Tsinghua University, Beijing 100084, China}
\affiliation{Department of Physics, Stanford University, Stanford, California 94305, USA}
\author{Fan Yang}
\email{yangfan\_blg@bit.edu.cn}
\affiliation{School of Physics, Beijing Institute of Technology, Beijing 100081, China}

\begin{abstract}
Whether the doped t-J model on the Kagome lattice supports exotic superconductivity has not been decisively answered. In this paper, we propose a new class of variational states for this model and perform large-scale variational Monte Carlo simulation on it. The proposed variational states are parameterized by the SU(2)-gauge-rotation angles, as the SU(2)-gauge structure hidden in the Gutzwiller-projected mean-field ansatz for the undoped model is broken upon doping. These variational doped states smoothly connect to the previously studied U(1) $\pi$-flux or $0$-flux states, and energy minimization among them yields a chiral noncentrosymmetric nematic superconducting state with $2 \times 2$-enlarged unit cell. Moreover, this pair density wave state possesses a finite Fermi surface for the Bogoliubov quasi particles. We further study experimentally relevant properties of this intriguing pairing state.
\end{abstract}
\date{\today}
\maketitle

{\bf Introduction:} Quantum spin liquids (QSL) have attracted increasing interest in condensed matter physics in the past decades \cite{Anderson73,rmp_qsl1,rmp_qsl2,rmp_qsl3,Broholm20,Balents10}. They represent an exotic class of insulating states which cannot be adiabatically connected into a trivial band insulator. Moreover, a QSL state can support fractionalized excitations with fractional braiding statistics. One of the most intriguing aspects of QSL lies in that doping a QSL might naturally lead to high temperature superconductivity\cite{Anderson87,Kivelson87,Rokhsar88,Laughlin88,Wen89,Wen96,Lee07,Fradkin15,Jiang2019,Jyf2020} or a topologically ordered Fermi liquid state (FL$^*$)\cite{Senthil03,Punk15,Patel16}.

One promising model exhibiting a QSL ground state is the spin-1/2 Heisenberg model on the Kagome lattice, which is probably realized by the spin-liquid candidate material Herbertsmithite\cite{rmp_qsl2}. Numerous efforts have been devoted to study properties of this model for several decades. Except for a few early results pointing toward the valence bond solid (VBS) state\cite{Huse1,Huse2,Vidal}, dominating numerical results suggest a QSL ground state for this model\cite{Jiang08,Yan11,Jiang12,Depenbrock12,Gong15,Mei17,Ran07,Iqbal13,Iqbal14,Liao17,He17,Taoli,Fradkin18}. Particularly, while a number of density-matrix renormalization group (DMRG) simulations on wide cylinders have exhibited evidences of a $Z_2$ QSL with exponentially decaying spin-spin correlation\cite{Jiang08,Yan11,Jiang12,Depenbrock12,Gong15,Mei17}, recent iDMRG simulation on infinite cylinders\cite{He17}, tensor-network simulation on infinite system\cite{Liao17}, and variational Monte Carlo (VMC) studies\cite{Ran07,Iqbal13,Iqbal14} suggest that the ground state is a gapless U(1) Dirac QSL with algebraic correlation. While further studies are still needed to reveal the precise nature of the ground state at half filling, it is also desired to study what quantum state would be obtained when mobile charge carriers are introduced into it by doping. Especially, can exotic superconductivity emerge upon doping the Kagome QSL state?

The nature of the lightly doped Kagome system described by the t-J model is not decisively known so far. Nonetheless, recent DMRG study on the model with moderate doping on the 4-leg cylinder provided convincing evidences of an insulating holon Wigner crystal\cite{Jiang17}. On the wider system, previous VMC investigation of this model on up to $8^2\times9$ lattice in certain doping range suggests that the $\pi$-flux Dirac U(1) spin liquid\cite{Ran07} is unstable against a 0-flux state with a VBC ordering\cite{Guertler11,Guertler13}. As the $\pi$-flux state has lower energy than the 0-flux state at half filling, it is obvious that the 0-flux state obtained by VMC at certain doping range cannot be continuously connected to the undoped $\pi$-flux QSL state\cite{Ran07}. It is natural to ask what is  the ground state for the lightly doped t-J model on the Kagome lattice assuming that the ground state of the undoped system is a U(1) Dirac QSL.

In this paper, we study the t-J model on the Kagome lattice in the very low doping regime which is expected to smoothly connect with U(1) spin liquid at half-filling\cite{Ran07} by performing VMC simulations. Our study is inspired by a crucial SU(2)-gauge structure\cite{Baskaran88,Affleck88,Dagotto88} hidden in the projective construction at half-filling: two different mean-field (MF) ansatzs related by an arbitrary local SU(2)-gauge rotation actually correspond to the same physical spin state after the Gutzwiller-projection. Such gauge-redundancy leads to a many-to-one labeling between the mean-field ansatzs and the projected wave function at half-filling\cite{Wen02}. At finite doping, the breaking of this gauge structure differentiates the many states related by the gauge-rotation, which form our variational groups. We choose the doped $0$-flux or $\pi$-flux states as our un-rotated starting points. Energy minimizations within both groups of variational states yield chiral noncentrosymmetric nematic superconducting states with $2\times2$-enlarged unit cell in the very low doping regime, with the gauge-rotated $\pi$-flux state smoothly connecting to the undoped $\pi$-flux QSL\cite{Ran07}. Remarkably, as the SU(2)-gauge rotation maintains the quasi-particle spectrum, the obtained superconducting states possess finite Fermi surface (FS) for the Bogoliubov quasi-particles. The physical properties of these pairing states are intriguing: although they are superconducting states, they resemble those of the normal FL in many aspects.

\begin{figure}[tb]
\centering
\includegraphics[width=\linewidth]{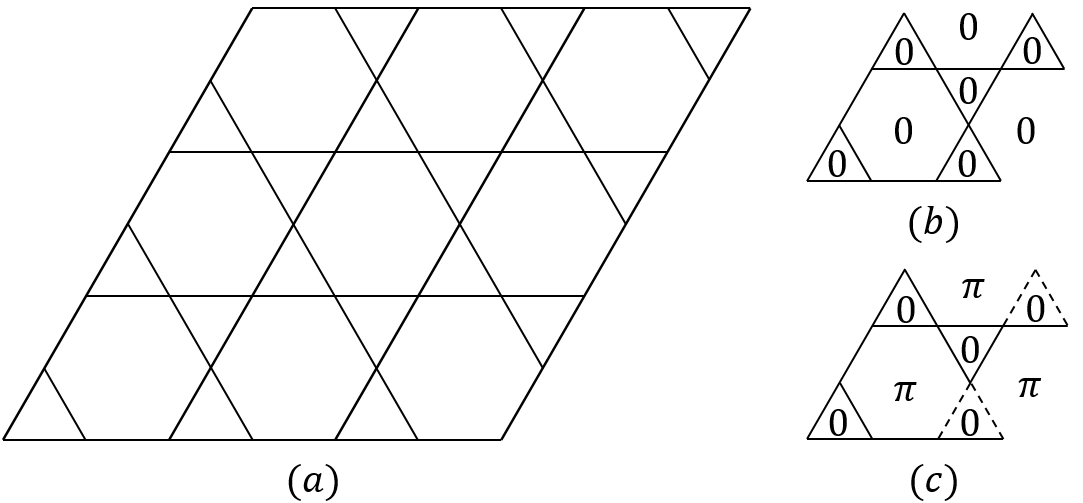}
\caption{(a) A schematic representation of the Kagome lattice. (b) The 0-flux state with $\chi_{ij}=1$ on each bond. (c) The $\pi$-flux state with zero flux through triangles and $\pi$-flux through hexagonals. Dashed lines indicate the $\chi=-1$ bonds. }
\label{piflux}
\end{figure}

{\bf Variational states:} We study the standard t-J model on the Kagome lattice illustrated in \Fig{piflux}(a):
\be
H= -t \sum_{\avg{ij}\sigma} P_G (c_{i\sigma}^\dagger c_{j\sigma} + h.c.)P_G + J\sum_{\avg{ij}} (\v{S}_i\cdot \v{S}_j-\frac{1}{4} n_i n_j),
\label{Ham}
\ee
where $c_{i\sigma}$ annihilates an electron on site $i$ with spin $\sigma$, $\v{S}_i=\frac12 c_{i\alpha}^\dagger \v{\sigma}_{\alpha\beta} c_{i\beta}$ denotes the spin operator and $n_i=\sum_{\sigma} c_{i\sigma}^\dagger c_{i\sigma} $ is the density operator. $P_G=\prod_i (1-n_{i\A} n_{i\V})$ is the Gutzwiller-projection operator enforcing no-double-occupancy constraint. $\avg{ij}$ represents nearest-neighbor (NN) bonding. Here we set $J=1$ as the energy scale. The parameter $t$ and the doping concentration $\delta$ are set as tuning parameters spanning the phase diagram.

To smoothly connect with the previously studied $\pi$-flux state at half-filling\cite{Ran07} and to compare energy with the zero-flux state at finite doping\cite{Guertler11,Guertler13}, we investigate the Gutzwiller-projected MF states generated by the following MF Hamiltonian,
\be
H_{MF}^0=\sum_{\avg{ij}\sigma} \chi_{ij} c_{i \sigma}^\dagger c_{j \sigma} + h.c.,
\ee
where $\chi_{ij}=\pm 1$. These states can be characterized by the fluxes $e^{i\phi}= \prod_{\rm plaquette} \sgn(\chi_{ij}) $ through triangle and hexagon plaquettes of the Kagome lattice. In this work, we primarily focus on two types of fluxes: (1) the 0-flux states having zero flux through all the triangles and hexagons shown in \Fig{piflux}(b); (2) the $\pi$-flux state having $\pi$ flux through the hexagons and zero flux through the triangles as shown in \Fig{piflux}(c). At half filling, both flux states after the projection are QSL. While the former has a large spinon FS, the latter is a U(1) Dirac QSL. Previous VMC studies\cite{Ran07} showed that the $\pi$-flux state has the lowest energy among all studied states.

The key point lying behind the present work is the following SU(2)-gauge structure hidden in the projective construction at half-filling\cite{Affleck88, Dagotto88}. Let's perform the following local SU(2)-gauge transformation $W_i$ on the two component spinor $\psi_i=(c_{i\uparrow},c^{\dagger}_{i\downarrow})^{T}$,
\bea
\begin{bmatrix}
c_{i\A} \\
c^\dagger_{i\V}
\end{bmatrix}
\rightarrow
W_i
\begin{bmatrix}
c_{i\A} \\
c^\dagger_{i\V}
\end{bmatrix}
.
\label{gauge_rotation}
\eea
At half-filling, any two MF ansatzs connected by this local SU(2)-gauge rotation label the same physical spin state after projected into the single-occupance subspace, as the spin operator $\v{S}_i$ keeps invariant under this gauge transformation\cite{Affleck88, Dagotto88}. However, this many-to-one labeling is absent once the system is doped away from half filling. Consequently, the many states related by the gauge rotation before projection can represent physical states with distinct physical properties at finite doping. One may naturally raise the following question: what is the lowest-energy state among all those gauge-rotated $\pi$- or $0$-flux states for the system with very low doping? To answer this question, we choose the local SU(2)-gauge rotation angles as variational parameters, from which we construct MF Hamiltonian to generate the variational physical states by projection, for energy minimization in both flux sectors.

Our trial wave functions are generated by the following local SU(2)-gauge-rotated Bogoliubov-de Genes (BdG) MF Hamiltonian,
\bea
H_{MF}&=&\sum_{ij}
\begin{bmatrix}
c^\dagger_{i\A} & c_{i\V}
\end{bmatrix}
W_i
\begin{bmatrix}
\chi_{ij} & 0 \\
0 & -\chi_{ji}
\end{bmatrix}
W^\dagger_j
\begin{bmatrix}
c_{j\A} \\
c^\dagger_{j\V}
\end{bmatrix}.
\label{mfstate}
\eea
Here the unrotated MF parameter $\chi_{ij}$ on the NN-bond $\avg{ij}$ for the $\pi$- and $0$-flux states have been introduced above. We set the on-site term $\chi_{ii}$ to a uniform value $\chi_{ii}=\chi_0$ as the chemical potential term. The local SU(2) rotation matrix $W_i$ can be parameterized by the following three rotation angles $\alpha_i,\beta_i$ and $\gamma_i$ as
\bea
W_i=
\begin{bmatrix}
e^{i \beta_i} \cos\alpha_i  & e^{i \gamma_i} \sin\alpha_i \\
-e^{-i \gamma_i} \sin\alpha_i  & e^{-i \beta_i} \cos\alpha_i
\end{bmatrix}.
\label{W_matrix}
\eea
Our trial wave function $P_G\ket{\Psi_{\text{MF}}\{\chi_0,\alpha,\beta,\gamma\}}$ now depends on the set of variational parameters $\{\alpha_i,\beta_i,\gamma_i\}_{i=1,\cdots,N}$ and $\chi_0$. Here $\ket{\Psi_{\text{MF}}\{\chi_0,\alpha,\beta,\gamma\}}$ is the MF ground state of Eq. (\ref{mfstate}).

{\bf VMC results:} We adopt standard Monte Carlo approach to simulate the variational states $P_G\ket{\Psi_{\text{MF}}\{\chi_0,\alpha,\beta,\gamma\}}$ on the Kagome lattice with size 3$\times L$$\times L$ and  periodic boundary condition, where the two adopted lattice sizes $L=8$ and $L=12$ lead to consistent results. The numerical complexity arising from optimizing a large number of variational parameters is overcome by the stochastic reconfiguration (SR) method \cite{Sorella05}. We further reduce the number of SU(2) rotation angles by restricting the parameters in the super-cell with size 3$\times 2$$\times 2$. We have checked that increasing the size of the super-cell does not lead to a lower optimized energy (See Supplemental material (SM) for detail).

\begin{figure}[tb]
\centering
\includegraphics[width=\linewidth]{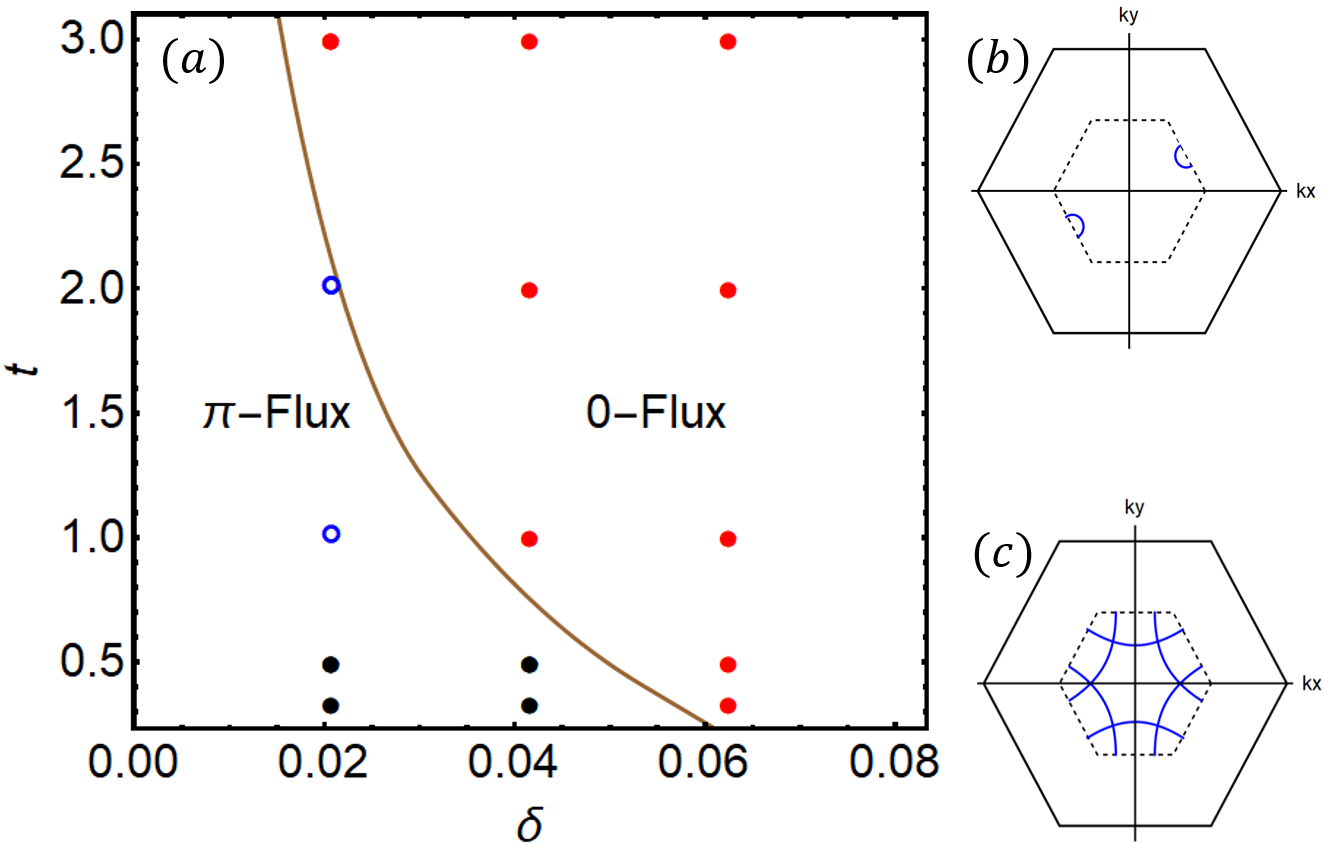}
\caption{(a) phase diagram of the slightly doped t-J model on a $8 \times 8 \times 3$ lattice. The black circles in the $\pi$-flux sector represent metallic phase without pairing. (b) Nearly doubly degenerate small FSs of the slightly doped $\pi$-flux state located around the two folded Dirac points of undoped state. (c) Folded FSs of the doped $0$-flux state. }
\label{phase}
\end{figure}

Our main results are summarized in the phase diagram shown in \Fig{phase}(a), where we consider several $t$ ranging from $1/3$ to $3$ and several doping levels below $7\%$  on the Kagome lattice with $L=8$. Starting from the undoped $\pi$-flux state, the lowest-energy state stays in the $\pi$-flux sector until beat by the optimized states in $0$-flux sector at a finite doping concentration $\delta_c$ depending on $t$. For small $t\sim 1/3$, the gauge-rotated $\pi$-flux state is stable until the doping concentration reaches $\delta_c\sim 5\%$. While for large $t$, a smaller doping is enough to drive the system away from the $\pi$-flux sector, consistent with previous VMC studies at $J=0.4t$ \cite{Guertler11,Guertler13}.
To explore the possible finite size effect, we also studied the models on $L=12$ lattice with 4 to 12 doped holes and find that the gauge-rotated $\pi$-flux state is still the lowest-energy state for most of the cases at small doping region.

The physical properties of the gauge-rotated $\pi$-flux phase are mainly determined by the optimized SU(2) rotation angles, which are provided in the SM\cite{SM}.
Except for the two parameter points in the small $J$ and $\delta$ region of the $\pi$-flux sector (black circles in \Fig{phase}), we find that the optimized angle $\alpha_i$ for both flux sectors are neither 0 nor $\pi$. Consequently, the non-zero off-diagonal terms in the gauge-rotation matrices $W_i$ defined in Eq. (\ref{W_matrix}) bring about a singlet pairing term $H_{\Delta}=-\sum_{ij}c^{\dagger}_{i\uparrow}c^{\dagger}_{j\downarrow}\left[\chi_{ij}e^{i\left(\beta_i+\gamma_j\right)}\cos\alpha_i\sin\alpha_j+(i\rightleftarrows j)\right]+h.c.$ in $H_{\text MF}$.  Note that the gauge rotation (\ref{gauge_rotation}) as a unitary transformation does not change the quasi-particle spectra \cite{Affleck88,Dagotto88}, but it only leads to enlargement of the unit cell. As a result, the superconducting states generated here will have quasi-particle FSs simply folded from those of the doped $0$- or $\pi$-flux states before the gauge rotation, as shown in \Fig{phase}(b) and (c). Therefore, we have obtained here singlet pairing states with finite Bogoliubov FS. Such SC states breaking translational symmetry with finite FS were pair-density-wave states\cite{Berg07,Tsunetsugu2008,Berg2009,Berg2009NP,Berg2010, Fradkin2012,Lee2014,Seamus2016,Seamus2019, YYWang2018,Yao2020,Yao2020prl,Yao2021,YXWang2020review}.

\begin{table}[tb]
\begin{tabular}{lcccccc}
\multicolumn{1}{l|}{\textbf{}}                 &  & $\delta=0.92\%$ & $\delta=1.85\%$ & $\delta=2.78\%$ \\
\hline
\multicolumn{1}{l|}{\multirow{5}{*}{t=2}}       & 0-flux     & -0.96037(2) & -1.00873(3) & -1.05669(3) \\
\multicolumn{1}{l|}{}                           & $\pi$-flux & -0.97105(5) & -1.01197(2) & -1.05238(3) \\
\multicolumn{1}{l|}{}                           & $Z_2$ QSL  & -0.97106(2) & -1.01196(4) & -1.05231(4) \\
\multicolumn{1}{l|}{}                           & VBC-D      & -0.96066(3) & -1.00921(2) & -1.05710(3) \\
\multicolumn{1}{l|}{}                           & CDW        & -0.9112(3) & / &  / \\
\hline
\multicolumn{1}{l|}{\multirow{5}{*}{t=1}}       & 0-flux     & -0.92894(3) & -0.94680(1) & -0.96408(2) \\
\multicolumn{1}{l|}{}                           & $\pi$-flux & -0.94347(2) & -0.95691(2) & -0.97010(4) \\
\multicolumn{1}{l|}{}                           & $Z_2$ QSL  & -0.94348(3) & -0.95686(4) & -0.97010(3) \\
\multicolumn{1}{l|}{}                           & VBC-D      & -0.92933(2) & -0.94698(2) & -0.96442(2) \\
\multicolumn{1}{l|}{}                           & CDW        & -0.9104(4) & / &  / \\
\hline
\multicolumn{1}{l|}{\multirow{5}{*}{t=0.5}}     & 0-flux     & -0.91336(4) & -0.91565(3) & -0.91772(3) \\
\multicolumn{1}{l|}{}                           & $\pi$-flux & -0.92967(3) & -0.92939(2) & -0.92828(2) \\
\multicolumn{1}{l|}{}                           & $Z_2$ QSL  & -0.92965(2) & -0.92936(3) & -0.92827(3) \\
\multicolumn{1}{l|}{}                           & VBC-D      & -0.91367(2) & -0.91588(3) & -0.91808(2) \\
\multicolumn{1}{l|}{}                           & CDW        & -0.9154(2) & / &  /
\end{tabular}
\caption{Optimized energy of part of the candidates on the model with $t=0.5\sim2$ and $\delta=0.92\%\sim 2.78\%$ on a $3 \times 12 \times 12$ lattice. A complete table with more candidate ansatz can be found in SM.}
\label{energy}
\end{table}

The optimized gauge-rotation angles in the $\pi$-flux sector are complicated because all the $\{ \alpha_i, \beta_i, \gamma_i \}$ within the super cell are non-zero and non-uniform, breaking the TRS, the lattice-rotation, the inversion and the translational symmetries.  The pairing and hopping terms generated by the gauge rotations are generally complex and are of the same order of magnitude, which suggests a typical inter-band pairing state. More details of the optimized gauge-rotation angles and the resulting gauge-rotated MF Hamiltonian are provided in the SM. In spite of the complicated pairing and hopping terms, the resulting MF Hamiltonian exhibits finite Bogoliubov FS shown in \Fig{phase}(b), which comprises two nearly doubly degenerate small pockets folded from those of the un-rotated $\pi$-flux state.

At infinitesimal doping, the gauge-rotated $\pi$-flux state is reasonably the lowest-energy VMC state due to the finite energy difference between this state and other states presented in the previous VMC study of the undoped case. When the doping concentration becomes larger, besides the gauge-rotated $\pi$- or $0$- flux states, other competitive states such as the holon Wigner crystal\cite{Jiang17}, the doped $Z_2$ QSL\cite{Taoli}, various types of VBC states\cite{Guertler11,Guertler13}, and the uniform-pairing states\cite{Anderson87,Gros88,Gutzwiller1} should also be considered in the VMC calculations. In Table \ref{energy}, we list the optimized energies for part of the lowest-energy states we obtained on the $L=12$ lattice, which suggests that in the small doping region the gauge-rotated $\pi$-flux state has lower energy than the other VMC candidates. We can see the doped $Z_2$ QSL \cite{Taoli} provides similar energy as the gauge-rotated $\pi$-flux state because after optimization such state actually flows back to the U(1) Dirac spin liquid ($\pi$-flux state) for all the cases we studied. In the $0$-flux sector, we find that the D-type VBC state has slightly lower energy than the gauge-rotated $0$-flux state. Another important candidate, the holon Wigner crystal, is mimicked by the CDW ansatz in the VMC calculation. Restricted by the finite lattice size, we only consider the four-hole doped $L=12$ system with $3 \times 6\times 6$ super-cell. Though we observe the similar density distribution as the Wigner crystal, the VMC energy of this CDW state is higher than the gauge-rotated $\pi$-flux state. We also consider the uniform-pairing states with both the extended s-wave and d-wave pairing parameters lived on the nearest and second nearest neighbor bonds, which also provide higher energies in the small doping region.
A more complete comparison of all the competing states we considered on the $L=12$ lattice and the detailed VMC realization of them are presented in the SM\cite{SM}.

\begin{figure}[tb]
\centering
\includegraphics[width=\linewidth]{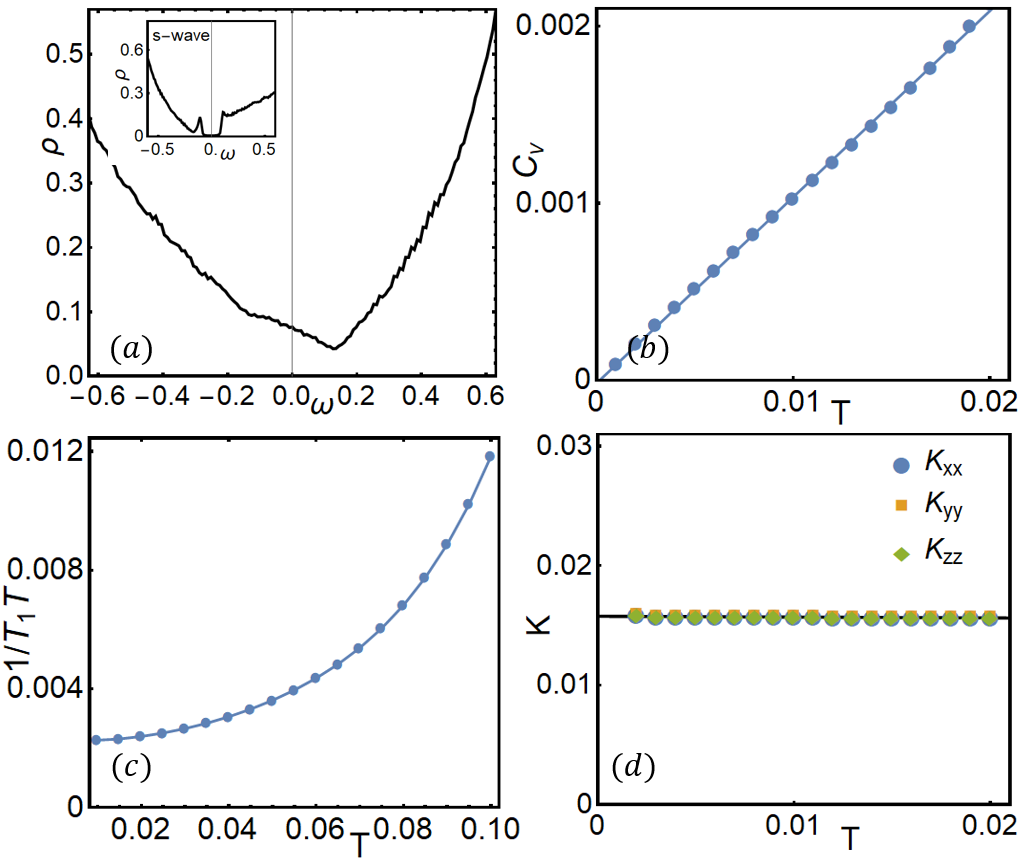}
\caption{Experiment-relevant quantities for the optimized gauge-rotated $\pi$-flux state. (a) $dI/dV\sim V$ curve for the STM. The inset is the $dI/dV$ curve for the model with uniform on-site s-wave (b) the specific heat $C_v\sim T$. (c) the NMR relaxation rate $1/T_1T$. (d) the NMR Knight-shift $K$ as function of $T$, three colors stand for $K_{xx}$, $K_{yy}$ and $K_{zz}$ respectively. The optimal gauge-rotation angles are obtained from parameter setting $t=0.5$ and $\delta=2.08\%$. }
\label{STM}
\end{figure}

{\bf Singlet pairing with finite FS:} The singlet pairing with Bogoliubov FS obtained here is distinct from conventional superconductors. To reveal the physical properties of this intriguing pairing state relevant to experiments, we shall perform MF studies below toward the zero- and finite-temperature behaviors of the system represented by the optimized $H_{MF}$. Consequently, this pairing state is found to be very exotic.

On one hand, the breaking of U(1)-gauge symmetry leads to finite superfluid density as expected (see SM\cite{SM} for details), which will result in detectable Meissner effect. On the other hand, the presence of the full FS causes finite density of state (DOS) which, in combination with the singlet-pairing signature, makes this pairing state look like a normal FL in the aspects of low lying quasi-particle and spin excitations, as shown in Fig. 3 for the gauge-rotated $\pi$-flux state. In the zero-temperature $dI/dV$ curve for the STM spectrum shown in Fig. 3(a), a finite zero-bias conductance appears caused by the finite DOS, in comparison with the U-shaped curve for the s-wave SC shown in the inset. Fig. 3(b) shows that the specific-heat $C_v\propto T$ at $T\to 0$, resembling the normal FL. Fig. 3(c) illustrates that the relaxation rate $1/T_1T$ of the nuclear magnetic resonance (NMR) saturates to a finite value at $T\to 0$, obeying a Korringa-law-like behavior for the FL, different from the $1/T_1T\to 0$ behavior for conventional fully-gapped ($\propto e^{-\Delta/T}$) or nodal ($\propto T^3$) SC. Fig. 3(d) exhibits that the NMR Knight-shift $K$ saturates to a finite value for $T \to 0$, independent of the orientation of the exerted magnetic field, similarly to the Pauli-susceptibility behavior for standard FL. This behavior is distinct from the $K\to 0$ behavior of conventional singlet SC with full or nodal gap or the obvious magnetic-field-orientation-dependence of $K$ for the triplet SC. 
Although both the gauge-rotated $\pi$- or $0$- flux states exhibit BFS, the different doping dependences of the area enclosed by their FSs can be distinguished by the ARPES, which can also lead to different behaviors such as the doping dependence of $C_v/T$. Details of these MF studies are provided in the SM\cite{SM}.

{\bf Discussion and Conclusion:} Note that, starting with a U(1) QSL at half-filling, we have only considered the gauge-rotation angles as variational parameters and neglect the amplitude fluctuation of $\chi_{ij}$ before the gauge rotation. Such a treatment is reasonable only at zero-doping limit. For higher dopings, lower variational energy is generally expected if we include the variation of the amplitude of $\chi_{ij}$. The band structure of such improved state can be strongly modified, i.e. Hastings-type VBC order can gap out the Dirac points\cite{Hastings00}. We have briefly investigated the fate of Hastings-type VBC in the unrotated $\pi$ flux state, and found that it becomes visible when the doping concentration is larger than $\delta_c \sim 4\%$. Therefore, close to the zero-doping limit, the Bogoliubov FS is more likely to survive. 

Previous studies\cite{Paramekanti2001,Paramekanti2004,Yunoki2005,Nave2006,Yang2007} have shown the survival of the FS under the Gutzwiller projection, although some other MF properties might be modified\cite{Ferrari2019}, such as the quasi-particle weight. Similar phenomenon, namely the survival of the FS under Gutzwiller projection, is also directly observed for our projected gauge-rotated states by numerically detecting the FS-jump in the occupation-number distribution of Bogoliubov quasiparticles in the momentum space (see the SM for details\cite{SM}). Another concern about the stability of the Bogoliubov FS obtained here under possible remnant interactions among the Bogoliubov quasi-particles neglected in the VMC treatment. Indeed the FSs shown in Fig.~\ref{phase}(b) and (c) satisfy the relation $\varepsilon_{\mathbf{k}}=\varepsilon_{\mathbf{-k}}$ as the unitary SU(2)-gauge rotation adopted here maintains the quasi-particle energy, which will suffer from the Cooper instability under remnant interactions. However, note that the two superconducting states obtained here break both the TRS and the inversion symmetry\cite{SM}. Without the protection of these two symmetries \cite{Yao2012}, the relation $\varepsilon_{\mathbf{k}}=\varepsilon_{\mathbf{-k}}$ cannot survive such perturbations as the further variations of $\{\chi_{ij},\Delta_{ij}\}$ after the gauge rotation, which can always exist for finite doping. Consequently, the Bogoliubov FSs obtained here should be stable against weak remnant interactions among the quasi-particles.

Evidences of SC with Bogoliubov FS can also appear in other contexts such as the FFLO state induced in the magnetic field\cite{FF,LO}, the cubic system with $j=3/2$ total-angular-momentum degree of freedom\cite{Agterberg2017} and some iron-based superconductors with spin-orbit coupling and interband pairing\cite{Setty2020}. The recently synthesized YPtBi multi-band superconductor with strong spin-orbit-coupling\cite{Brydon2016,Kim2018} might also exhibits Bogoliubov FS if it breaks the TRS\cite{Timm2017}. While these systems host similar normal FL-like quasi-particle excitations as here, their spin excitations have different properties from those of the singlet pairing state obtained here. In summary, we propose a new way to obtain the Bogoliubov FS: doping a U(1) QSL. The key point lies in that the local SU(2)-gauge rotation, which brings about SC to the doped QSL, will not alert the quasi-particle energy, which is different from doping a QSL with spinon FS \cite{PALee2019}. Such mechanism not only applies to the doped Kagome U(1) QSL, but also applies to other doped U(1) QSL, which could be a promising way to obtain the new type of unconventional gapless SC in strongly-correlated electronic systems.

{\it Acknowledgment: }We are grateful to the helpful discussions with T. Li, Y.-M. Lu, Y. Zhou, W.-Q. Chen, Z.-C. Gu and Z.-Y. Weng. This work is supported in part by the Department of Energy, Office of Science, Basic Energy Sciences, Materials Sciences and Engineering Division, under Contract DE-AC02-76SF00515 (YFJ), the NSFC Grants No. 11674025 (FY), 11825404 (HY), the MOSTC under Grant Nos. 2016YFA0301001 and 2018YFA0305604 (HY), the Strategic Priority Research Program of Chinese Academy of Sciences under Grant No. XDB28000000 (HY), the Beijing Municipal Science \& Technology Commission under grant No. Z181100004218001 (HY), the Beijing Natural Science Foundation under grant No. Z180010 (HY). HY would also like to acknowledge support in part by the Gordon and Betty Moore Foundations EPiQS Initiative through Grant GBMF4302. Parts of the computing for this work was performed on the Sherlock cluster.

\renewcommand{\theequation}{A\arabic{equation}}
\setcounter{equation}{0}
\renewcommand{\thefigure}{A\arabic{figure}}
\setcounter{figure}{0}
\renewcommand{\thetable}{A\arabic{table}}
\setcounter{table}{0}
\begin{widetext}
\section{Supplement Material}
\subsection{Rotated Hamiltonian}
After SU(2) rotation $\{ W_i \}$ defined in main text, the new Hamiltonian still has the compact form
\bea
H_{MF}=
\begin{bmatrix}
c^\dagger_{\A} & c_{\V}
\end{bmatrix}
\begin{bmatrix}
\chi & \Delta \\
\Delta^\dagger & -\chi^T
\end{bmatrix}
\begin{bmatrix}
c_{\A} \\
c^\dagger_{\V}
\end{bmatrix}
\label{mfstate_a} ,
\eea
where $c_{\A}=\{c_{1\A}, c_{2\A}, \dots , c_{N\A}\}^T$ and the matrix elements of $\chi$ and $\Delta$ regarding the $i$ and $j$ sites now take more complicated forms:
\bea
\left(
\begin{array}{cc}
 \chi_0  \cos 2 \text{$\alpha_i$} & \chi_{ij}(\sin \text{$\alpha_i$} \sin \text{$\alpha_j$} e^{i
   (\text{$\gamma_i$}-\text{$\gamma_j$})}-  \cos \text{$\alpha_i$} \cos \text{$\alpha_j$} e^{-i
   (\text{$\beta_i$}-\text{$\beta_j$})}) \\
 \chi_{ij}  (\sin \text{$\alpha_i$} \sin \text{$\alpha_j$} e^{-i(\text{$\gamma_i$}-\text{$\gamma
  _j$})}-\cos \text{$\alpha_i$} \cos \text{$\alpha_j$} e^{i(\text{$\beta_i$}-\text{$\beta
  _j$})}) & \chi_0  \cos 2 \text{$\alpha_j$} \\
\end{array}
\right) ,
\label{kinetic}
\eea
and
\bea
\left(
\begin{array}{cc}
 \chi_0 e^{-i (\text{$\beta_i$}-\text{$\gamma_i$})} \sin 2\text{$\alpha_i$} & -\chi_{ij} \left(\sin \text{$\alpha_i$} \cos \text{$\alpha_j$}
   e^{i (\text{$\gamma_i$}-\text{$\beta_j$})}+\cos \text{$\alpha_i$} \sin \text{$\alpha_j$} e^{i
   (\text{$\gamma_j$}-\text{$\beta_i$})}\right) \\
 -\chi_{ij} \left(\sin \text{$\alpha_i$} \cos
   \text{$\alpha_j$} e^{i (\text{$\gamma_i$}-\text{$\beta_j$})}+\cos \text{$\alpha_i$} \sin
   \text{$\alpha_j$} e^{i (\text{$\gamma_j$}-\text{$\beta_i$})}\right) & \chi_0 e^{-i (\text{$\beta_j$}-\text{$\gamma_j$})} \sin 2 \text{$\alpha_j$}  \\
\end{array}
\right).
\label{pairing}
\eea

\begin{figure}[tbh]
\centering
\includegraphics[height=4.0cm]{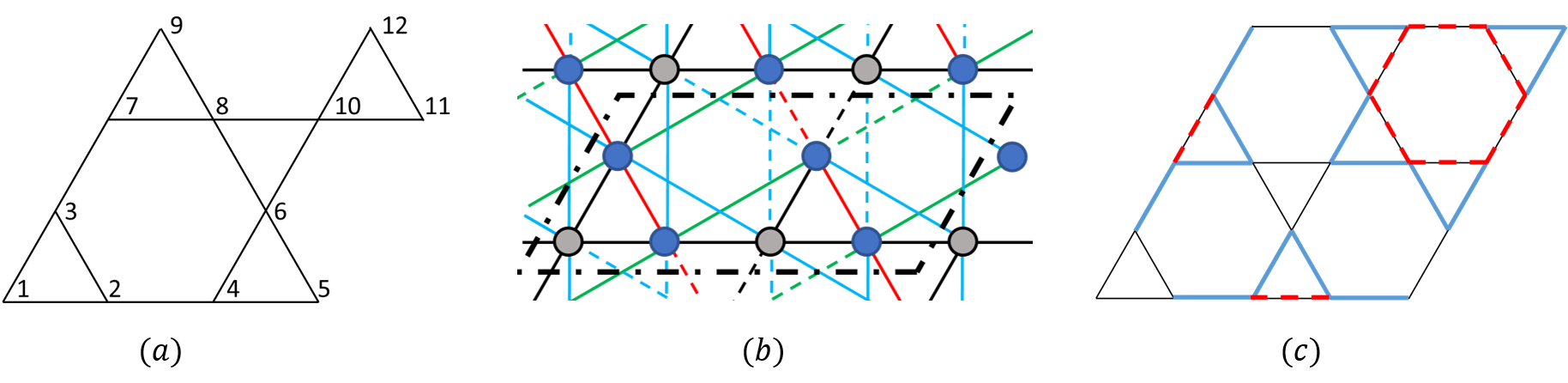}
\caption{(a) site labels of the enlarged 12-site unit cell. (b) The illustration of $Z_2$ ansatz for doped $Z_2$ QSL, the color of bond and sites  are explained in the text.  (c) The illustration of the D-type VBC ansatz. }
\label{sites}
\end{figure}

\subsection{Comparing with other candidates}


\begin{table}[tb]
\begin{tabular}{l|ccc|ccc|ccc|}
               &  \multicolumn{3}{c|}{ $\delta=0.92\%$} & \multicolumn{3}{c|}{ $\delta=1.85\%$} &  \multicolumn{3}{c|}{ $\delta=2.78\%$}  \\
               & t=0.5  &  t=1  &   t=2  &  t=0.5  & t=1  &  t=2  & t=0.5  & t=1  &  t=2  \\
\hline
0-flux + rot.      & -0.91336(4) & -0.92894(3) & -0.96037(2) & -0.91565(3) & -0.94680(1) & -1.00873(3)  & -0.91772(3) & -0.96408(2) & -1.05669(3) \\
\hline
$\pi$-flux + rot.  & -0.92967(3) & -0.94347(2) & -0.97105(5) & -0.92939(2) & -0.95691(2) & -1.01197(2)  & -0.92828(2) & -0.97010(4) & -1.05238(3) \\
\hline
$Z_2$ QSL          & -0.92965(2) & -0.94348(3) & -0.97106(2) & -0.92936(3) & -0.95686(4) & -1.01196(4)  & -0.92827(3) & -0.97010(3) & -1.05231(4)  \\
\hline
VBC-D              & -0.91367(2) & -0.92933(2) & -0.96066(3) & -0.91588(3) & -0.94698(2) & -1.00921(2) & -0.91808(2) & -0.96442(2) & -1.05710(3) \\
\hline
$0$-flux+CDW       &  -0.8968(7) & -0.8949(5) &  -0.8951(6)  &  / & /  & / & / & / & /  \\
\hline
$\pi$-flux+CDW     &  -0.9154(2) & -0.9104(4) & -0.9112(3) &  / & /  & / & / & / & /  \\
\hline
$0$-flux+SC        & -0.90851(2) & -0.92397(2) & -0.95549(3) & -0.9103(1) & -0.9411(2) & -1.0038(1)  & -0.9125(1) & -0.9596(1) & -1.0519(4) \\
\hline
$\pi$-flux+SC      & -0.90671(3) & -0.90752(5) & -0.9078(1) & -0.8891(2) & -0.8937(1) & -1.0003(5)  & -0.9130(3) & -0.9574(1) & -1.0492(1)  \\
\hline
\end{tabular}
\caption{The optimized energy of several candidate states of Kagome t-J model obtained on $3\times 12 \times 12$ lattice with different doping concentration $\delta$. The first two lines are the energy of 0- and $\pi$-flux states supplemental by an additional $SU(2)$ rotations. $Z_2$ in the third line stands for doped $Z_2$ QSL ansatz. VBC-D means the doped D-type VBC. CDW stands for our ansatz \Eq{wigner} which mimics the holon Wigner crystal state. SC in the last two column means the general superconducting states with pairing up to next nearest neighbor bond.}
\label{energyall}
\end{table}

\emph{Doped $Z_2$ QSL: }
At zero doping limit, the $Z_2$ QSL is known to be one of the most competitive ground-state candidate of the Kagome Heisenberg model. In lightly doped region, it is therefore important to check whether the $Z_2$ QSL ansatz can become the lowest-energy state of the corresponding t-J model. To answer this question, we investigate the optimized state of the extended $Z_2$ ansatz via VMC and compare the energy with the results of our U(1) QSL ansatz. We start from the original undoped $Z_2$ ansatz suggested in Ref.[34], which is defined as
\bea
H_{MF}&=&\sum_{ij}
\begin{bmatrix}
c^\dagger_{i\A} & c_{i\V}
\end{bmatrix}
U_{ij}
\begin{bmatrix}
c_{j\A} \\
c^\dagger_{j\V}
\end{bmatrix}.
\label{mfstate_Z2}
\eea
where $U_{ij}=\begin{bmatrix}
\chi_{ij} & \Delta_{ij} \\
\Delta^\dagger_{ij} & -\chi_{ji}
\end{bmatrix}$ is a $2\times 2$ matrix with matrix elements $\chi_{ij}$ and $\Delta_{ij}$ representing hopping and pairing parameters for the ansatz. We used the same form of $U_{ij}$ defined in Ref.[34], which are illustrated in \Fig{sites}(b) as
\bea
U_{ii}=
\begin{cases}
\mu \tau_3 & \text{gray sites} \\
\mu \vec{n}_{\phi_1}\cdot \vec{\tau} & \text{navy sites}
\end{cases}, \
U_{ij}=-s_{ij}
\begin{cases}
\tau_3 & \text{black bonds} \\
\vec{n}_{\phi_2}\cdot \vec{\tau} & \text{red bonds}
\end{cases}, \
U_{ij}=-\nu_{ij}
\begin{cases}
\eta \vec{n}_{\phi_3}\cdot \vec{\tau} & \text{blue bonds} \\
\eta \vec{n}_{\phi_4}\cdot \vec{\tau} & \text{green bonds}
\end{cases}.
\eea
Here, $\vec{\tau}=(\tau_1, \tau_2, \tau_3)$ are the Pauli matrixes, $\vec{n}_\phi=(\sin(\phi),0,\cos(\phi))$. $\mu$ and $\eta$ are two real number. $s_{ij}=\pm 1$ on the solid(dashed) NN bonds and $\nu_{ij}=\pm 1$ on the solid(dashed) NNN bonds in \Fig{sites}(b). This $Z_2$ ansatz also has the SU(2)-gauge structure in the projective construction at half filling. To make a fair comparison with the U(1) case, we similarily treat this local gauge-rotation as an additional set of order parameters and the extended $Z_2$ ansatz now becomes to
\bea
H_{MF}&=&\sum_{ij}
\begin{bmatrix}
c^\dagger_{i\A} & c_{i\V}
\end{bmatrix}
W_i U_{ij} W^\dagger_j
\begin{bmatrix}
c_{j\A} \\
c^\dagger_{j\V}
\end{bmatrix},
\label{Z2state}
\eea
where $W_i$ is the local gauge-transformation defined in the Eq.(3) in the main text. The optimized energy of the extended $Z_2$ ansatz on the $L=12$ lattice is listed in Table \ref{energyall}, where we can see that its energy is the same as that of the $\pi$-flux state within statistic error for all the cases we tried. The reason of this can be explained by looking at the optimized variational parameters of the $Z_2$ ansatz. In all the lightly doped cases we studied, the optimized values of the U(1)-symmetry-breaking variational parameters $\phi_1 \sim \phi_4$ and $\eta$ converge to vanishing values $<10^{-3}$. Within the convergence accuracy of these variational parameters determined by the statistic error of the energy calculation, these variational parameters can be viewed as zero, indicating that the state actually reduces to the doped U(1) $\pi$-flux state. Note that these nearly vanishing values make the convergence of the doped $Z_2$ QSL slower than that of the $\pi$-flux state, for most of the cases more iterations of the SR method is required for the doped $Z_2$ QSL to obtain the equally converged energy.

\emph{Holon Wigner crystal:}
One of the important competitive ground state candidate of the doped QSL on the Kagome lattice is the holon Wigner crystal reported in the previous DMRG study[36]. On the long cylinder, DMRG study finds an insulating charge density wave with one doped hole per enlarged unit cell. Though the lattice geometry and boundary condition in the DMRG study is different from our VMC study, the key features of the density profile of holon Wigner crystal can be mimic by the VMC ansatz with the enlarged super-cell and site dependent chemical potentials, which leads to the mean field Hamiltonian
\bea
H_{MF}=\sum_{\avg{ij}\sigma} \chi_{ij} c_{i \sigma}^\dagger c_{j \sigma} + h.c. - \sum_i \mu_i n_i,
\label{wigner}
\eea
where $\mu_i$ is the site-dependent chemical potential and $\chi_{ij}$ is the hopping parameter. To ensure the number of holes in each enlarged unit cell are exactly one, we select the size of enlarged super-cell such that the number of the unit cell equals to the number of doped holes. For example, we can divide the lattice into four $3 \times 6 \times 6$ super-cells for a $L=12$ system with 4 doped holes and treat the hopping parameters and 108 chemical potentials $\mu_i$ as variational parameters in VMC calculation. Since both spin and SC correlation functions in the Wigner crystal are short-range, here we do not introduce any other spin and SC order parameter into the ansatz.

The optimization is applied by the following two steps. First, we fix the hopping $\chi_{ij}=\pm 1$ of the correspinding 0- or $\pi$-flux state and consider only $\mu_i$  as variational parameters. At this step, we find the hole quickly localized on a single site of the supercell and density profile becomes similar to the one obtained in the DMRG study. Then we relax the hole by including the hopping parameters $\chi_{ij}$ around the hole as additional variational parameters. The optimized energy will slightly decrease during this precedure but no remarkable changes is observed in density profiles for both 0-flux and $\pi$-flux state. The final energy of this CDW ansatz is listed in Table \ref{energyall}, where we can see that energy of this simplified ansatz is higher than the gauge-rotated $\pi$-flux state. 

\emph{Valence bond crystal: }
Previous study focusing on slightly larger doping concentration[37,38] reported that the valence bond crystal with zero flux has the lowest energy. To compare the energy of VBC candidate state with the one obtained from new $\pi$-flux state, we check the optimized energy of the improved Hasting type VBC, D-type VBC, in the small doping region. The ansatz of the D-type VBC can be written as
\bea
H_{MF}=\sum_{\avg{ij}\sigma} \chi_{ij} c_{i \sigma}^\dagger c_{j \sigma} + h.c. ,
\eea
here the $\chi_{ij}$ is $\chi_1$ on the blue bond in \Fig{sites}(c), $\chi_2$ on the red dashed bond and 1 on all the rest bond. As listed in Table \ref{D-VBC}, we find that the optimized $\chi_1$ and $\chi_2$ at light doping is nearly one for all the cases we studied.

\begin{table}[tbh]
\begin{tabular}{|l|ccc|ccc|ccc|}
\hline
               &  \multicolumn{3}{c|}{ $\delta=0.92\%$} & \multicolumn{3}{c|}{ $\delta=1.85\%$} &  \multicolumn{3}{c|}{ $\delta=2.78\%$}  \\
                         & t=0.5  &  t=1  &   t=2  &  t=0.5  & t=1  &  t=2  & t=0.5  & t=1  &  t=2  \\
\hline
$\chi_1$            & 1.006  & 1.008 & 1.006 &  1.02  &  1.02  &  1.02  & 1.03 & 1.03 & 1.03  \\
\hline
$\chi_2$            & 0.995  & 0.994 & 0.996 &  0.99  &  0.99  &  0.97  & 0.96 & 0.96 & 0.96  \\
\hline
\end{tabular}
\caption{The optimized $\chi_{ij}$ for the D-type VBC of a series of systems on $3\times 12\times 12$ lattice.}
\label{D-VBC}
\end{table}

\emph{Uniform pairing states: }
Here we study more conventional Gutzwiller-projected BCS-MF states with uniform NN-bond pairings grown on top of the doped $0$- and $\pi$- flux states. For singlet pairing, the general variational Hamiltonian can be written as
\bea
H_{MF}&=&\sum_{ij}
\begin{bmatrix}
c^\dagger_{i\A} & c_{i\V}
\end{bmatrix}
\begin{bmatrix}
\chi_{ij} & \Delta_{ij} \\
\Delta^\dagger_{ij} & -\chi_{ji}
\end{bmatrix}
\begin{bmatrix}
c_{j\A} \\
c^\dagger_{j\V}
\end{bmatrix}
\label{SM:mfstate}
\eea
where $\chi_{ij}$ on each bonds follow the pattern shown in Fig. 1(b) or (c), onsite $\chi_{ii}=\mu$ is the chemical potential adjusting the average electron number. Pairing $\Delta_{ij}=\Delta^r_s+\Delta^r_d e^{2i\theta_{ij}}$ are non-zero only on the nearest neighbor (NN) and next rest neighbor (NNN) bond. Here $r$ stands for the NN and NNN bond, $\Delta^r_s$ represents the extended $s$-wave pairing and $\Delta^r_d$ is the strength of the $d$-wave pairing with $\theta_{ij}$ denoting the azimuth of $\mathbf{r}_j-\mathbf{r}_i$ which can be  $0$, $\pi/3$ and $2\pi/3$ depending on the directions of the bonds. We ignore the onsite $s$-wave pairing as it will be projected out after Gutzwiller projection. We then treat all the $\Delta$ as complex numbers to include relative phases between pairing channels and take $\{\Delta^{NN}_s, \Delta^{NN}_d, \Delta^{NNN}_s, \Delta^{NNN}_d, \mu\}$ as variational parameter to calculate the optimized energy of $0$ and $\pi$-sector. In TABLE \ref{energyall}, we compare their energy, labeled by  $0$-flux+SC and  $\pi$-flux+SC respectively, with the energy of other canditate. For small doping concentration $\delta\le 2.16\%$, the energy obtained from the $s$-wave and $d$-wave pairing state are all higher than the lowest energy of the non-trivial SC state in the main text. Similar result is also found on the $L=8$ lattice.

\subsection{Optimized angles}
As listed in Table \ref{Oangles}, we select several typical optimized angles obtained from different points of the phase diagram: $\pi$-flux state at $t=1$ and $\delta=1.04\%$ (left panel); $\pi$-flux state at $t=0.5$ and $\delta=2.08\%$ (middle panel); $0$-flux state at $t=1$ and $\delta=2.08\%$ (right panel). The left panel is an example of metal phase in $\pi$-flux sector consists of nearly zero $\alpha$ angles and non-zero $\beta$ angle as phases of the new hopping terms. The $\gamma$ term in this case is negligible because of the vanishing off-diagonal term. The middle panel shows a superconducting state in the $\pi$-flux sector. The right panel exhibits the optimized angle obtained in the $0$-flux sector with vanishing $\gamma$ and uniform non-zero $\alpha$ angles indicating the superconducting nature of this phase.

\begin{table}[h]
\begin{tabular}{cccc}
\multicolumn{1}{c|}{i}  &  $\alpha$ & $\beta$ & $\gamma$ \\
\hline
\multicolumn{1}{c|}{1}  & -0.008 &  2.027 &  -1.815 \\
\multicolumn{1}{c|}{2}  &  0.007 &  2.839 &   0.895 \\
\multicolumn{1}{c|}{3}  & -0.008 &  1.338 &  -1.443 \\
\multicolumn{1}{c|}{4}  &  0.008 & -2.561 &   0.443 \\
\multicolumn{1}{c|}{5}  &  0.008 &  2.453 &   0.897 \\
\multicolumn{1}{c|}{6}  &  0.007 &  0.341 & -3.140 \\
\multicolumn{1}{c|}{7}  &  0.007 &  1.021 &   2.041 \\
\multicolumn{1}{c|}{8}  & -0.008 &  0.642 &  -0.756 \\
\multicolumn{1}{c|}{9}  &  0.008 &  0.646 &   1.693 \\
\multicolumn{1}{c|}{10} &  0.008 &  1.235 &   2.736 \\
\multicolumn{1}{c|}{11} & -0.008 & -3.078 &  -0.757 \\
\multicolumn{1}{c|}{12} & -0.007 &  0.247 &   3.141\\
\end{tabular}~~~~~~~~
\begin{tabular}{cccc}
\multicolumn{1}{c|}{i}  &  $\alpha$ & $\beta$ & $\gamma$ \\
\hline
\multicolumn{1}{c|}{1}  & -1.085 & -2.811 &  2.193 \\
\multicolumn{1}{c|}{2}  &  2.319 &  0.755 & -0.764 \\
\multicolumn{1}{c|}{3}  & -0.905 &  1.293 &  2.015 \\
\multicolumn{1}{c|}{4}  & -0.670 & -3.111 &  2.648 \\
\multicolumn{1}{c|}{5}  & -2.321 &  0.120 &  2.388 \\
\multicolumn{1}{c|}{6}  &  0.429 &  2.727 & -3.138 \\
\multicolumn{1}{c|}{7}  & -0.808 & -3.036 &  1.801 \\
\multicolumn{1}{c|}{8}  & -0.531 & -2.896 &  1.460 \\
\multicolumn{1}{c|}{9}  & -2.252 &  1.958 &  2.018 \\
\multicolumn{1}{c|}{10} &  0.421 & -1.857 & -2.300 \\
\multicolumn{1}{c|}{11} & -2.602 &  0.666 &  1.477 \\
\multicolumn{1}{c|}{12} &  2.745 &  2.222 &  0.000 \\
\end{tabular}~~~~~~~~
\begin{tabular}{cccc}
\multicolumn{1}{c|}{i}  &  $\alpha$ & $\beta$ & $\gamma$ \\
\hline
\multicolumn{1}{c|}{1}  & -1.158 & -0.443 &   0.0000 \\
\multicolumn{1}{c|}{2}  & -1.143 & -0.629 &  -0.0002 \\
\multicolumn{1}{c|}{3}  & -1.141 &  1.407 &  -0.0001 \\
\multicolumn{1}{c|}{4}  & -1.160 & -0.573 &  -0.0001 \\
\multicolumn{1}{c|}{5}  & -1.141 & -0.534 &  -0.0002 \\
\multicolumn{1}{c|}{6}  & -1.146 &  0.529 &   0.0000 \\
\multicolumn{1}{c|}{7}  & -1.159 & -0.346 &   0.0001 \\
\multicolumn{1}{c|}{8}  & -1.145 & -1.808 &   0.0002 \\
\multicolumn{1}{c|}{9}  & -1.137 & -0.979 &   0.0002 \\
\multicolumn{1}{c|}{10} & -1.163 & -0.542 &   0.0003 \\
\multicolumn{1}{c|}{11} & -1.150 & -0.240 &   0.0002 \\
\multicolumn{1}{c|}{12} & -1.149 &  1.810 &   0.0000 \\
\end{tabular}
\caption{Optimized rotation angles obtained from different points in the phase diagram of $L=8$ lattice. From left to right: $\pi$-flux state at $t=1$ and $\delta=1.04\%$; $\pi$-flux state at $t=0.5$ and $\delta=2.08\%$; $0$-flux state at $t=1$ and $\delta=2.08\%$.}
\label{Oangles}
\end{table}

Because the optimized angle generally break lattice-rotation, inversion and translational symmetry, the SC breaking translational symmetry are expected. As an concrete example, we measure the pairing order parameter of the $\pi$-flux state at $t=0.5$ and $\delta=2.08\%$, e.g., singlet SC order on the translational related bonds $(1,2)$, $(4,5)$, $(7,8)$,$(10,11)$ illustrated in \Fig{sites} are $0.017+0.199i$, $0.166+0.144i$, $-0.017+0.214i$ and $-0.178+0.034i$ respectively. We also measure the density profile of the same model which breaks the translational symmetry, e.g., the projected charge density on sites 1, 4, 7 and 11 are 0.933, 0.960, 0.950 and 0.976 respectively.

\subsection{Experiment-related quantities}
Here we study the experiment-related properties of optimized $H_{\text MF}$ with full FS. We perform the MF studies on the models at both zero and finite temperature with $200\times 200$ 12-site unit cells. In the $\pi$-flux sector, the optimized angles shown in the middle panel of Table \ref{Oangles} are used as a representative point.
The specific heat of the system is given by
\bea
C_v=\frac{1}{2N}\sum_{\v{k}n}E_{\v{k}n} \frac{d f(E_{\v{k}n})}{dT}
\eea
where $f$ is the Fermi distribution, N denotes the total number of lattice site. $E_{\v{k}n}$ is the energy of the rotated Hamiltonian $H_{\text MF}$ where $\v{k}$ and $n$ label momentum and index of eigenvalues respectively.

The STM spectrum can be written as
\bea
\rho_{\mu}(\omega)=\frac{1}{N}\text{Im} \sum_{\v{k}n} \frac{\abs{\braket{\v{k}\mu}{\v{k}n}}^2}{\omega-E_{\v{k}n}-i 0^+} ,
\eea
where $\mu$ labels the original band index of the model. In practice, the $0^+$ is replaced by the small interval $\Delta \omega$.

The Knight shift is proportional to the spin susceptibility $\chi_{ss}$, which gives
\bea
K_{ss}\propto -\frac{1}{N}\sum_{\v{k}mn}\abs{\bra{\v{k}m}S_s\ket{\v{k}n}}^2 \frac{f(E_{\v{k}m})-f(E_{\v{k}n})}{E_{\v{k}m}-E_{\v{k}n}}
\eea
where $S_s$ is the matrix of $s$ component of the spin operator in the Nambu space.

The NMR spin-relaxation rate reads
\bea
\frac{1}{T_1 T} \propto -\frac{1}{N^2}\sum_{\v{k}\v{k'}mns}A(\v{k'}-\v{k})\abs{\bra{\v{k}m}S_s\ket{\v{k'}n}}^2\frac{\partial f(E)}{\partial E}|_{E=E_{\v{k}m}}\delta(E_{\v{k}m}-E_{\v{k'}n}) .
\eea
Here for simplicity we set the geometrical structure factor $A(\v{q})$ to $1$ and replace the delta function by a Lorentzian, $\delta(E) \rightarrow \frac{\Gamma}{\pi(E^2+\Gamma^2)}$. Due to the heavy computational cost of relaxation rate we reduce the lattice size from $200\times 200$ to $100\times 100$ and increase the temperature interval in this calculation.

\begin{figure}[tb]
\centering
\includegraphics[width=0.6\linewidth]{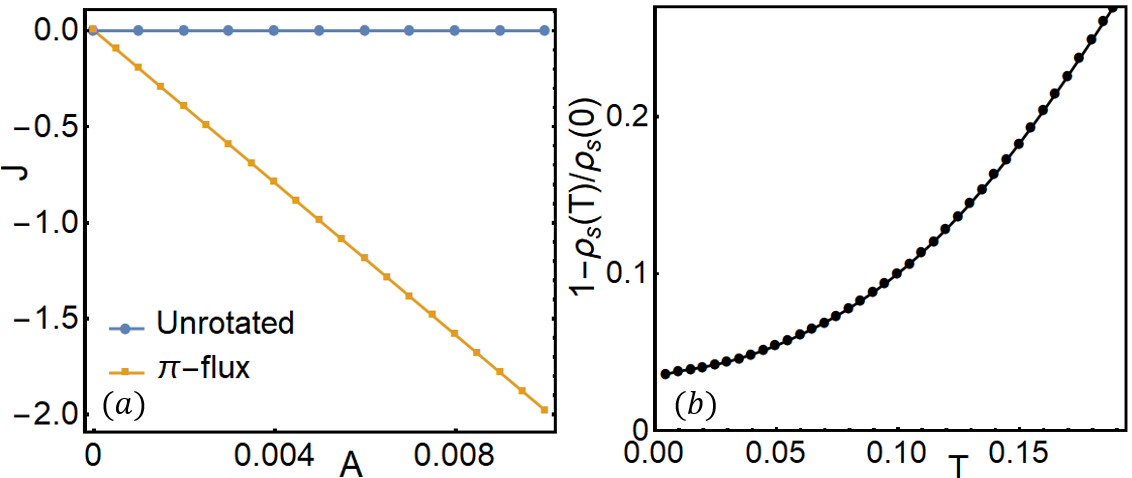}
\caption{Superfluid density of the SU(2)-gauge rotated $\pi$-flux state for (a) zero- and (b) finite- temperatures. The Superfluid density $\rho_s$ is obtained from the slope of induced current $\v{j}_s(\v{A})=-\rho_s \v {A}$. The same parameter set as the one in Fig. 3 of main text is used.}
\label{meissner}
\end{figure}

The current operator $\v{j}_i$ at site $i$ is defined as $\v{j}_i=\frac{\partial H_{\text MF}^k(\v{A})}{ \partial \v{A}}$, where $H_{\text MF}^k(\v{A})=\sum_{ij} e^{i \int_i^j \v{A}\cdot dl} h_{ij}c^\dagger_i c_j$ is obtained from the kinetic part of the rotated Hamiltonian $H_{\text MF}^k=\sum_{ij} h_{ij}c^\dagger_i c_j$ expressed in \Eq{kinetic}. Here spin is omitted for simplicity. In the weak $A$ limit, up to $O(A)$ order, we have
\bea
\v{j}_i=\sum_{j} \frac{h_{ij}}{2}(-i+\v{A}_i\cdot \v{R}_{i \-> j})\v{R}_{i \-> j} c^\dagger_i c_j + h.c.
\eea
where vector $\v{R}_{i\->j}$ points from site $i$ to site $j$. Because of $h_{ij}$, the current operator strongly depends on the SU(2) rotation $\{ W_i \}$ defined in main text. By solving the ground-state of the mean-field Hamiltonian $H_{\text MF} (\v{A})$, we can numerically obtain the superfuild density $\rho_s$ from fit  $\v{j}_s=-\rho_s \v{A}$ (setting $e=m=1$) shown in \Fig{meissner}. As known for the PDW with Fermi surface, the zero-temperature $\v{j}_s\sim\v{A}$ relation of the SU(2)-gauge rotated $\pi$-flux state shown in \Fig{meissner}(a) exhibits a finite negative slope, different from the zero slope for the normal unrotated state.
The temperature dependence of $1-\rho_s(T)/\rho_s(0)$ is shown in \Fig{meissner}(b), which exhibits a jump at $T=0$ caused by the many quasi-particle excitations across the FS.

\subsection{Larger super-cell}
We exam several different sizes of the enlarged super cell to check how the optimized energy depends on the choice of super-cell. As shown in the Table.\ref{supercell}, the energy of the $\pi$-flux state does not decrease when the size is larger than $3\times 2 \times 2$ for $L=12$ system, indicate that the $3\times 2 \times 2$ super-cell is large enough for this study.
\begin{table}[tbh]
\begin{tabular}{|l|c|c|c|c|c|c|}
\hline
Size      & $3 \times 2 \times 1$  & $3 \times 2 \times 2$ &  $3 \times 2 \times 3$ & $3 \times 4 \times 4$  \\
\hline
Energy    &  -0.9239         &             -0.9297         &            -0.9296     &  -0.9297      \\
\hline
\end{tabular}
\caption{The optimized energy of the $\pi$-flux state on $L=12$, $t=0.5$ and $4$ hole doped system with varies size of super-cell.}
\label{supercell}
\end{table}

\subsection{Projected Bogoliubov Fermi surface}
Here we present a direct numerical evidence of the effect of the Gutzwiller projection on the Bogoliubov FS. For this purpose, we launch a MC calculation on a large lattice with $N=12\times 12\times 12=1728$ sites (here the first 12 stands for the site number of the $2\times 2$ enlarged super cell of our state) to obtain the k-space distribution of the occupation numbers of the Bogoliubov quasiparticles, i.e.
\begin{equation}
n_{\mathbf k\sigma}^{\gamma}\equiv \sum_{\alpha=11}^{12}\left\langle \gamma^{\dagger}_{\mathbf k\alpha\sigma}\gamma_{\mathbf k\alpha\sigma}\right\rangle
\end{equation}
in the folded Brilloiun zone in the obtained Gutzwiller-projected gauge-rotated $\pi$-flux state. Here the $\gamma_{\mathbf k\alpha\sigma}$ denotes the annihilation operator of the Bogoliubov quasi-particle with momentum $\mathbf k$, spin $\sigma$ and the band index $\alpha$. Note that the summation includes the two bands $\alpha=11 \sim 12$ which cross the FS. Here the slightly large doping concentration $\delta=7.17\%$ is adopted to ensure that the FS is big enough to be observed without any ambiguity. The variational parameters are fixed by energy minimization through our VMC calculation on a smaller lattice with $N=3\times 12\times 12=432$ sites with $t=0.5$, $J=1$ and close doping level in the $\pi$-flux sector.

\begin{figure}[tbh]
  \centering
  \includegraphics[height=4.0cm]{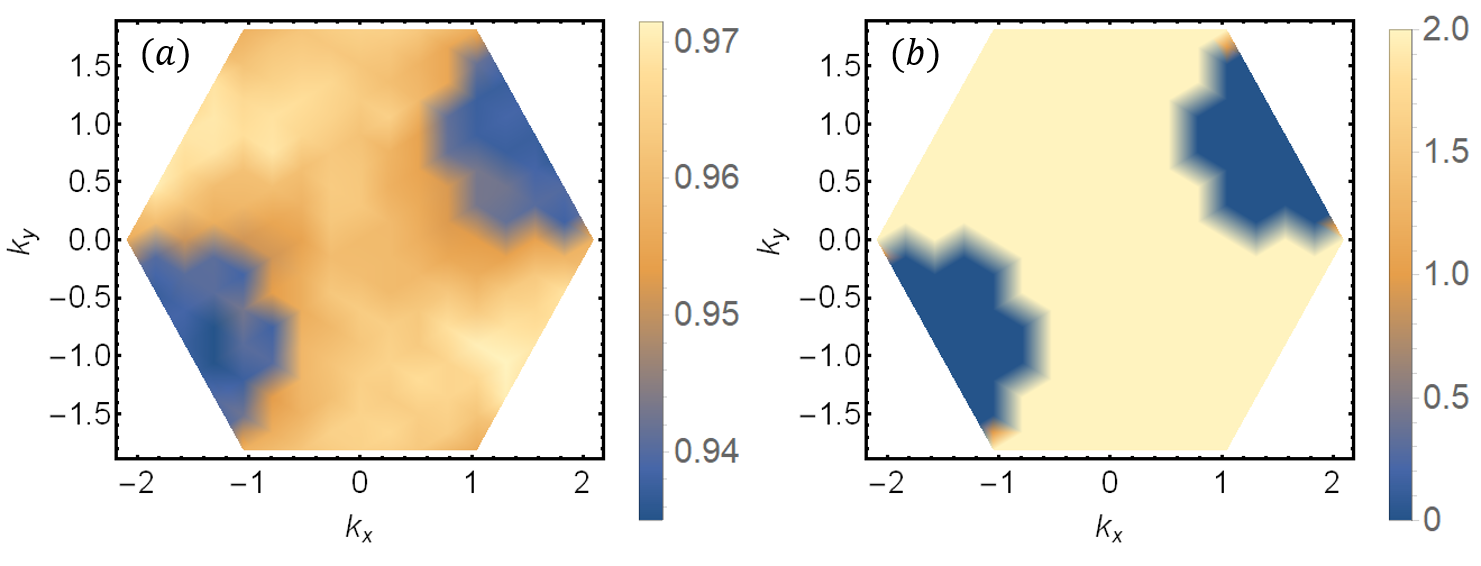}
  \caption{ (a) The k-space distribution of the occupation numbers of the Bogoliubov quasiparticles per spin in the folded Brillouin zone for the Gutzwiller-projected gauge-rotated $\pi$-flux state on a lattice with $12 \times 12 \times 12$ sites. (b) Result for the un-projected MF state on the same lattice. The colors in both panels represent the occupation numbers. The maximum occupation number per spin is 2 because of the additional two-fold degeneracy caused by the band folding.} 
  \label{fig_bfs}
\end{figure}

The numerical result is shown in the \Fig{fig_bfs}. For both the Gutzwiller-projected and un-projected MF states, we find the clear FS jumps in the distributions of the occupation numbers of the Bogoliubov quasiparticles within the folded Brillouin zone, suggesting the existence of well-defined FSs in both cases. Comparing the \Fig{fig_bfs}(a) and (b), one finds that the FSs of both states coincide with each other, and the main effect of the Gutzwiller projection lies in the renormalization of the jumped value across the FS, or equally the quasi-particle weight, by a factor at the order of doping $\delta$. These results are qualitatively consistent with the slave-boson-mean-field theory.

\end{widetext}


\begin{thebibliography}{99}
\bibitem{Anderson73} P.W. Anderson, Mater. Res. Bull. {\bf 8}, 153 (1973).
\bibitem{rmp_qsl1} Y. Zhou, K. Kanoda, and T.-K. Ng, Rev. Mod. Phys. {\bf 89}, 025003 (2017).
\bibitem{rmp_qsl2} M. R. Norman, Rev. Mod. Phys. {\bf 88}, 041002 (2016).
\bibitem{rmp_qsl3} P. A. Lee, N. Nagaosa, and X.-G. Wen, Rev. Mod. Phys. {\bf 78}, 17 (2006).
\bibitem{Broholm20} C. Broholm, R. J. Cava, S. A. Kivelson, D. G. Nocera, M. R. Norman, and T. Senthil, Science {\bf 367}, eaay0668 (2020).


\bibitem{Balents10} L. Balents, Nature (London) {\bf 464}, 199 (2010).

\bibitem{Anderson87} P. W. Anderson, Science {\bf 235}, 1196 (1987).
\bibitem{Kivelson87} S. A. Kivelson, D. S. Rokhsar, and J. P. Sethna, Phys. Rev. B {\bf 35}, 8865 (1987).
\bibitem{Rokhsar88} D. S. Rokhsar and S. A. Kivelson, Phys. Rev. Lett. {\bf 61}, 2376 (1988).
\bibitem{Laughlin88} R. B. Laughlin, Science {\bf 242}, 525 (1988).
\bibitem{Wen89} X. G. Wen, F. Wilczek, and A. Zee, Phys. Rev. B {\bf 39},11413 (1989).
\bibitem{Wen96} X.-G. Wen and P. A. Lee, Phys. Rev. Lett. {\bf 76}, 503 (1996).
\bibitem{Lee07} S. S. Lee,  P. A. Lee,   and T. Senthil, Phys. Rev. Lett. {\bf 98}, 1 (2007).
\bibitem{Fradkin15} E. Fradkin, S. A. Kivelson, and J. M. Tranquada, Rev. Mod. Phys. {\bf 87}, 457 (2015).
\bibitem{Jiang2019} H.-C. Jiang, arXiv:1912.06624 (2019).
\bibitem{Jyf2020} Y.-F. Jiang and H.-C. Jiang, arXiv:2002.04686 (2020).

\bibitem{Senthil03} T. Senthil, S. Sachdev, and M. Vojta, Phys. Rev. Lett. {\bf 90}, 216403 (2003).
\bibitem{Punk15} M. Punk, A. Allais, and S. Sachdev, PNAS {\bf 112}, 9552 (2015).
\bibitem{Patel16} A. A. Patel, D. Chowdhury, A. Allais, and S. Sachdev, Phys. Rev. B {\bf 93}, 165139 (2016).

\bibitem{Huse1} R. R. P. Singh and D. A. Huse, Phys. Rev. B {\bf 76}, 180407(R) (2007).
\bibitem{Huse2} R. R. P. Singh and D. A. Huse, Phys. Rev. B {\bf 77}, 144415 (2008).
\bibitem{Vidal} G. Evenbly and G. Vidal, Phys. Rev. Lett. {\bf 104}, 187203 (2010).


\bibitem{Jiang08} H. C. Jiang, Z. Y. Weng, and D. N. Sheng, Phys. Rev. Lett. {\bf 101}, 117203 (2008).
\bibitem{Yan11} S. Yan, D. Huse, and S. White, Science {\bf 332}, 1173 (2011).
\bibitem{Jiang12} H. C. Jiang, Z. Wang, and L. Balents, Nature Physics {\bf 8}, 902 (2012).
\bibitem{Depenbrock12} S. Depenbrock, I. P. McCulloch, and U. Schollw{\"o}ck, Phys. Rev. Lett. {\bf 109}, 067201 (2012).
\bibitem{Gong15} S.-S. Gong, W. Zhu, L. Balents, and D. N. Sheng, Phys. Rev. B {\bf 91}, 075112 (2015).
\bibitem{Mei17} J.-W. Mei, J.-Y. Chen, H. He, and X.-G. Wen, Phys. Rev. B, {\bf 95}, 235107 (2017).

\bibitem{He17} Y.-C. He, M. P. Zaletel, M. Oshikawa, and F. Pollmann, Phys. Rev. X  {\bf 7}, 031020 (2017).
\bibitem{Liao17} H. J. Liao, Z. Y. Xie, J. Chen, Z. Y. Liu, H. D. Xie, R. Z. Huang, B. Normand, and T. Xiang, Phys. Rev. Lett. {\bf 118}, 137202 (2017).
\bibitem{Ran07} Y. Ran, M. Hermele, P. A. Lee, and X. G. Wen, Phys. Rev. Lett. {\bf 98}, 117205 (2007).
\bibitem{Iqbal13} Y. Iqbal, F. Becca, S. Sorella, and D. Poilblanc, Phys. Rev. B {\bf 87}, 060405 (2013).
\bibitem{Iqbal14} Y. Iqbal, D. Poilblanc, and F. Becca, Phys. Rev. B {\bf 89}, 020407 (2014).
\bibitem{Taoli} T. Li, arXiv:1807.09463.
\bibitem{Fradkin18} H. J. Changlani, D. Kochkov, K. Kumar, B. K. Clark, and E. Fradkin, Phys. Rev. Lett. {\bf 120}, 117202 (2018).

\bibitem{Jiang17} H.-C. Jiang, T. Devereaux, and S. A. Kivelson, Phys. Rev. Lett. {\bf 119}, 067002 (2017).

\bibitem{Guertler11} S. Guertler and H. Monien, Phys. Rev. B {\bf 84}, 174409 (2011).
\bibitem{Guertler13} S. Guertler and H. Monien, Phys. Rev. Lett. {\bf 111}, 097204 (2013).

\bibitem{Baskaran88} G. Baskaran and P. W. Anderson, Phys. Rev. B {\bf 37}, 580(R) (1988).
\bibitem{Affleck88} I. Affleck, Z. Zou, T. Hsu, and P. W. Anderson, Phys. Rev. B {\bf 38}, 745 (1988).
\bibitem{Dagotto88} E. Dagotto, E. Fradkin, and A. Moreo, Phys. Rev. B {\bf 38}, 2926 (1988).

\bibitem{Wen02} X.-G. Wen, Phys. Rev. B {\bf 65}, 165113 (2002).

\bibitem{Sorella05} S. Sorella, Phys. Rev. B {\bf 71}, 241103(R) (2005).

\bibitem{SM} See the Supplementary Material at http:\.......  for the formula of the SU(2)-gauge rotated mean-field Hamiltonian; the realization and optimized energy of  the holon Wigner crystal, the doped $Z_2$ QSL, various types of VBC states, and the uniform-pairing states; the optimized results of the SU(2)-gauge-rotation angles for the doped $\pi$- or $0$- flux states; the formula for the calculations of the STM, the specific heat, the Knight-shift, the NMR relaxation rate, the zero- and finite-temperature superfluid density.

\bibitem{Berg07} E. Berg, E. Fradkin, E.-A. Kim, S. A. Kivelson, V. Oganesyan, J. M. Tranquada, and S. C. Zhang, Phys. Rev. Lett. {\bf 99}, 127003 (2007).
\bibitem{Tsunetsugu2008} D. F. Agterberg and H. Tsunetsugu, Nature Physics {\bf 4}, 639 (2008).
\bibitem{Berg2009NP} E. Berg, E. Fradkin and S. A. Kivelson, Nature Physics {\bf 5}, 830 (2009).
\bibitem{Berg2009} E. Berg, E. Fradkin, S. A. Kivelson, and J. M. Tranquada, New J. Phys. {\bf 11}, 115004 (2009).
\bibitem{Berg2010} E. Berg, E. Fradkin, and S. A. Kivelson, Phys. Rev. Lett. {\bf 105}, 146403 (2010).
\bibitem{Fradkin2012} A. Jaefari and E. Fradkin, 
Phys. Rev. B {\bf 85}, 035104 (2012).
\bibitem{Lee2014} P. A. Lee, Phys. Rev. X {\bf 4}, 031017 (2014).


\bibitem{Seamus2016} M. H. Hamidian, S. D. Edkins, S. H. Joo, A. Kostin, H. Eisaki, S. Uchida, M. J. Lawler, E.-A. Kim, A. P. Mackenzie, K. Fujita, J. Lee and J. C. Seamus Davis, Nature {\bf 532}, 343 (2016).
\bibitem{YYWang2018} W. Ruan, X. Li, C. Hu, Z. Hao, H. Li, P. Cai, X. Zhou, D.-H. Lee and Y. Wang, Nature Physics {\bf 14}, 1178 (2018).
\bibitem{Seamus2019} S. D. Edkins, A. Kostin, K. Fujita, A. P. Mackenzie, H. Eisaki, S. Uchida, S. Sachdev, M. J. Lawler, E.-A. Kim, J. C. Seamus Davis, and M. H. Hamidian, Science {\bf 364}, 976 (2019).

\bibitem{Yao2020} S.-K. Jian, M. M. Scherer, and H. Yao, Phys. Rev. Research {\bf 2}, 013034 (2020).

\bibitem{Yao2020prl} Z. Han, S. A. Kivelson, and H. Yao, Phys. Rev. Lett. 125, 167001 (2020).
\bibitem{Yao2021} K. S. Huang, Z. Han, S. A. Kivelson, and H. Yao, arXiv:2103.04984.
\bibitem{YXWang2020review} D. F. Agterberg, J. S. Davis, S. D. Edkins, E. Fradkin, D. J. Van Harlingen, S. A. Kivelson, P. A. Lee, L. Radzihovsky, J. M. Tranquada, and Y. Wang, Annu. Rev. Condens. Matter Phys. 11, 231 (2020).


\bibitem{Gros88} C. Gros, Phys. Rev. B {\bf 38}, 931(1988).

\bibitem{Gutzwiller1} P. W. Anderson, M. Randeria, T. Rice, N. Trivedi, and F. Zhang, J. Phys. Cond. Matter {\bf 16}, R755 (2004).

\bibitem{Hastings00} M. B. Hastings, Phys. Rev. B {\bf 63}, 014413 (2000).

\bibitem{Yao2012} M. Barkeshli, H. Yao, and S. A. Kivelson, Phys. Rev. B {\bf 87}, 140402(R) (2013).

\bibitem{Paramekanti2001} A. Paramekanti, M. Randeria and N. Trivedi, Phys. Rev. Lett. {\bf 87}, 217002 (2001).
\bibitem{Paramekanti2004} A. Paramekanti, M. Randeria and N. Trivedi, Phys. Rev. B {\bf 70}, 054504 (2004).
\bibitem{Yunoki2005} S. Yunoki, Phys. Rev. B {\bf 72}, 092505 (2005).
\bibitem{Nave2006} C. P. Nave, D. A. Ivanov, and P. A. Lee, Phys. Rev. B {\bf 73}, 104502 (2006).
\bibitem{Yang2007} H.-Y. Yang, F. Yang, Y.-J. Jiang, and T. Li, Journal of Physics: Condensed Matter, 19 (2007).

\bibitem{Ferrari2019} F. Ferrari and F. Becca, Phys. Rev. X {\bf 9}, 031026 (2019).

\bibitem{FF} P. Fulde and R. A. Ferrell, Phys. Rev. {\bf 135}, A550 (1964).
\bibitem{LO} A. I. Larkin and Y. N. Ovchinnikov, Zh. Eksp. Teor. Fiz. {\bf 47}, 1136 (1964).

\bibitem{Agterberg2017} D. F. Agterberg, P. M. R. Brydon and C. Timm, Phys. Rev. Lett. {\bf 118}, 127001 (2017).
\bibitem{Setty2020} C. Setty, S. Bhattacharyya, Y. Cao, A. Kreisel and P. J. Hirschfeld, Nat. Comm. {\bf 11}, 523 (2020).

\bibitem{Brydon2016} P. M. R. Brydon, L. M. Wang, M. Weinert, and D. F. Agterberg, Phys. Rev. Lett. {\bf 116}, 177001 (2016).
\bibitem{Kim2018} H. Kim, K. Wang, Y. Nakajima, R Hu, S. Ziemak, P. Syers, L Wang, H. Hodovanets, J. D. Denlinger, P. M. R. Brydon, D. F. Agterberg, M. A. Tanatar, R. Prozorov and J. Paglione, Science Advances {\bf 4}, 4 (2018).
\bibitem{Timm2017} C. Timm, A. P. Schnyder, D. F. Agterberg and P. M. R. Brydon, Phys. Rev. B {\bf 96}, 094526 (2017).

\bibitem{PALee2019} X. Y. Xu, K. T. Law, and Patrick A. Lee, Phys. Rev. Lett. {\bf 122}, 167001 (2019).


\end{thebibliography}
\end{document}